\documentclass[twocolumn,prl,aps,showpacs,superscriptaddress]{revtex4}

\usepackage{graphicx}

\def\prb{Phys. Rev. B }
\def\prl{Phys. Rev. Lett. }

\def\be{\begin{equation}}
\def\ee{\end{equation}}
\def\ba{\begin{eqnarray}}
\def\ea{\end{eqnarray}}

\def\LSCO{La$_{2-x}$Sr$_x$CuO$_4$ }

\def\124{YBa$_2$Cu$_4$O$_8$ }

\def\C60{A$_x$C$_{60}$ }

\def\LBCO{La$_{2-x}$Ba$_{x}$CuO$_4$ }

\begin{document}

\title{Enhancement of the superconducting transition temperature in La$_{2-x}$Sr$_{x}$CuO$_{4}$ bilayers:
Role of pairing and phase stiffness}

\author{Ofer Yuli}
\affiliation{Racah Institute of Physics, The Hebrew University of
Jerusalem, Jerusalem 91904, Israel}

\author{Itay Asulin}
\affiliation{Racah Institute of Physics, The Hebrew University of
Jerusalem, Jerusalem 91904, Israel}

\author{Leonid Iomin}
\affiliation{Department of Physics, Technion - Israel Institute of
Technology, Haifa 32000, Israel}

\author{Gad Koren}
\affiliation{Department of Physics, Technion - Israel Institute of
Technology, Haifa 32000, Israel}

\author{Oded Millo}
\email{milode@vms.huji.ac.il} \affiliation{Racah Institute of
Physics, The Hebrew University of Jerusalem, Jerusalem 91904,
Israel}

\author{Dror Orgad}
\email{orgad@phys.huji.ac.il} \affiliation{Racah Institute of
Physics, The Hebrew University of Jerusalem, Jerusalem 91904, Israel}

\begin{abstract}

The superconducting transition temperature, $T_c$, of bilayers
comprising underdoped La$_{2-x}$Sr$_{x}$CuO$_{4}$ films capped by a thin heavily
overdoped metallic La$_{1.65}$Sr$_{0.35}$CuO$_4$ layer, is found
to increase with respect to $T_c$ of the bare underdoped films.
The highest $T_c$ is achieved for $x=0.12$, close to the
'anomalous' 1/8 doping level, and exceeds that of the
optimally-doped bare film.
Our data suggest that the enhanced superconductivity is confined
to the interface between the layers. We attribute the effect to a
combination of the high pairing scale in the underdoped layer with
an enhanced phase stiffness induced by the overdoped film.
\end{abstract}


\maketitle

There is considerable evidence that $T_c$ in the underdoped (UD)
regime of the cuprate high-temperature superconductors is governed
by phase fluctuations while some sort of pairing occurs at
considerably higher temperatures
\cite{Uemura,Kivelson,Corson,Ong-Nernst,ourreview,Raman}, akin to
the case of granular superconductors \cite{Merchant}. In contrast,
the overdoped (OD) region is more conventional in the sense that
pairing and phase order take place simultaneously. Consequently,
systems which are composed of layers of UD and OD cuprates
constitute a unique laboratory for studying the interplay between
superconductivity's two necessary ingredients: pairing and phase
coherence. Such systems may also serve as models of the naturally
occurring multi-layered cuprate compounds, such as the Hg-series,
where measurements of the $^{63}$Cu Knight shift have demonstrated
that in every unit cell the outer planes tend to become OD, while
the inner planes become UD \cite{Tokunaga,Kotegawa2}. From a
practical point of view, the UD-OD multilayers offer the enticing
prospect of raising $T_c$ above that of both components, by
combining the high pairing scale of the UD layers with the large
phase stiffness of the OD layers \cite{Kivelson2,Berg}.

\begin{figure}
\includegraphics[width=3.1in]{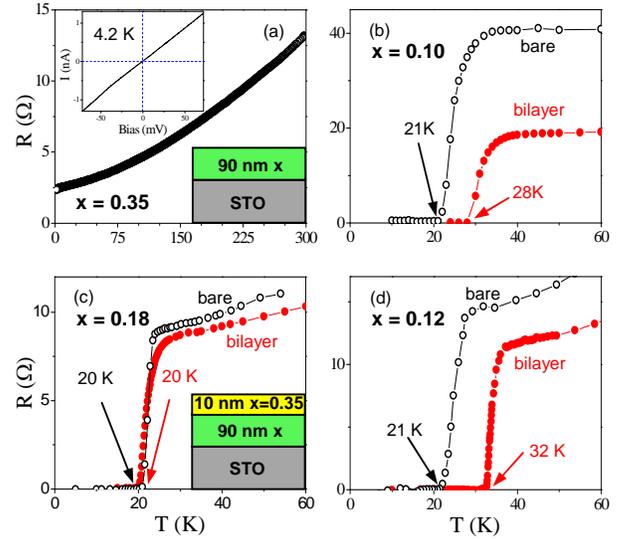}
\caption{(color online). (a) $R(T)$ of a bare LSCO($0.35$)
film. The inset depicts the $I-V$ tunneling characteristic of the same film,
taken by STM at 4.2~K . (b-d) $R(T)$ curves of the LSCO(0.35)$\,$--$\,$LSCO($x$)
bilayers with $x=0.10, 0.18$ and $0.12$, and of the corresponding bare films.
The arrows mark the zero-resistance transition temperature.}
\label{fig1}
\end{figure}

In this letter we present a systematic study of
La$_{1.65}$Sr$_{0.35}$CuO$_4$$\;$--$\;$\LSCO
[LSCO(0.35)$\;$--$\;$LSCO($x$)] bilayers, where $x$ varies from
the UD to the OD regime. Our most significant finding is an
enhancement of $T_c$ in bilayers containing an UD ($x < 0.15$)
layer. The highest $T_c$, well above that of the optimally-doped
bare film, was achieved for bilayers with $x = 0.12$, close to the
'anomalous' $x=1/8$ doping level. $T_c$ did not change when the
bottom layer was overdoped. Our magnetization measurements,
tunneling spectra, temperature-dependent resistance data and
non-linear $V(I)$ characteristics suggest that the enhanced
superconductivity occurs at the interface between the layers. We
attribute the $T_c$ enhancement (beyond strain effects that cannot
fully account for our observations), to an effective combination
of the high pairing scale of the UD layer with an increased phase
stiffness at the interface, induced by pair-propagation through
the OD component. We also point out that the fact that the maximal
$T_c$ enhancement occurs at $x = 0.12$ may reflect on the role of
stripes in the high-temperature superconductors.

LSCO($x$) films and LSCO(0.35)$\,$--$\,$LSCO($x$) bilayers with $x
= 0.06, 0.08, 0.10, 0.12$ (UD), $x = 0.15$ (optimally doped) and
$x = 0.18$ (OD) were epitaxially grown on (100) SrTiO$_3$ (STO)
wafers by laser ablation deposition [see schematic illustration in
Figs. 1(a) and 1(c)]. The LSCO($x$) films were 90~nm thick, and
the LSCO(0.35) overlayer, grown \emph{in situ} without breaking
the vacuum, was 10~nm thick. X-ray measurements confirmed a
\emph{c}-axis orientation perpendicular to the substrate. Temperature-dependent resistance, $R(T)$, measurements
were performed using the standard 4-probe technique. Special care
was taken to stabilize the temperature before each resistance
measurement and to avoid sample heating. We have also measured the
properties of a bare 90~nm LSCO(0.35) film, grown on STO, as
presented in Fig. 1(a). The $R(T)$ data showed no sign of a
superconducting transition down to a temperature of 2~K. Tunneling
spectra taken at 4.2~K using a scanning tunneling microscope (STM) exhibited Ohmic behavior; see
inset to Fig. 1(a). Therefore, we conclude that the $x=0.35$ layer
is metallic in the temperature range of our experiments ($T > 4.2$
K).

\begin{figure}
\includegraphics[width=\linewidth]{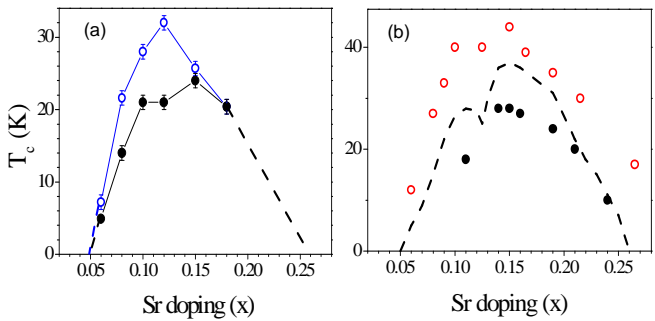}
\caption{(color online). (a) $T_c$ vs. $x$ of the bilayers (open symbols)
and bare films (solid symbols) measured in this work.
(b) $T_c$ of LSCO films grown on LaSrAlO$_4$ (open symbols), and
on STO (solid symbols), as compiled from Refs. \onlinecite{Sato}
and \onlinecite{Sato2}. The dotted line depicts the $T_c$ of bulk
LSCO.}\label{fig2}
\end{figure}

\begin{figure}
\includegraphics[width=1.8in]{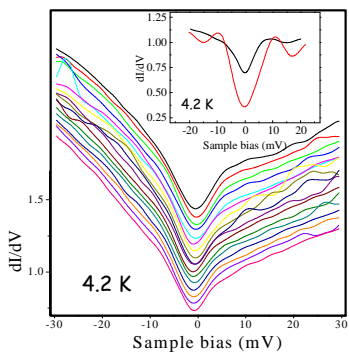}
\caption{(color online). Tunneling spectra of a bilayer composed of a 5~nm LSCO(0.35)
film on top of a 90~nm LSCO(0.10) layer. The data
were taken at 4.2~K and at equidistant steps along
a 31~nm long line. Inset: A spectrum of the bare
LSCO(0.10) film (red curve) and of the bilayer (black curve).}
\label{fig3}
\end{figure}

Typical $R(T)$ curves of $x = 0.10, 0.12$ and $0.18$ bilayers are
presented in Fig. 1(b-d) along with the corresponding bare film
data. The relatively low $T_c$ values of the bare films agree with
previous studies of LSCO films grown on STO
\cite{Kao,Sato,Curras}, as also shown in Fig. 2(b). The tensile
strain generated by the lattice-constant mismatch between the film
and substrate causes the transition temperature to drop below the
$T_c$ of the corresponding bulk sample. As illustrated in Fig.
1(b), the transition temperature of an
LSCO(0.35)$\,$--$\,$LSCO(0.10) bilayer was higher than the $T_c$
of the LSCO(0.10) UD bare film. On the other hand, no effect on
$T_c$ was observed for bilayers with an LSCO(0.18) OD film, see
Fig. 1(c). This contrast in the behavior of the UD and OD bilayers
is clearly apparent in Fig. 2(a), which presents a compilation of
the zero-resistance transition temperatures of all the films and
bilayers measured by us. An enhancement of $T_c$ was observed for
all the UD bilayers studied, with a magnitude that decreased both
towards the UD boundary of the superconducting phase and the
optimal doping level of the bare films. Surprisingly, the largest
enhancement, of 11~K, was found for the $x = 0.12$ bilayer.
Moreover, the $T_c$ of the LSCO(0.35)$\,$--$\,$LSCO(0.12) bilayer
was higher than those of both the bare optimally-doped LSCO(0.15)
film and its bilayer. We have also measured a sequence of inverted
bilayers, where a 10~nm LSCO($x$) film was deposited on top of a
90~nm LSCO(0.35) layer, and have found essentially the same
behavior. It is also worth noting that, in a control experiment,
no $T_c$ enhancement was observed in bilayers of gold and
LSCO(0.12).

Establishment of bulk superconductivity is accompanied by a diamagnetic
Meissner signal. With our SQUID magnetometer sensitivity we could not detect
any such signal at the enhanced $T_c$ of our bilayers. However, a clear
diamagnetic response was observed when each bilayer was cooled through the
transition temperature of the corresponding bare LSCO($x$) film.
This behavior points to the fact that the enhancement
does not occur in the bulk of the sample, but is likely an interface
phenomenon. We find further support for this conclusion in our STM data.

The tunneling spectra of our bilayers, measured by an STM on the
surface of the LSCO(0.35) top layer, exhibited a predominantly
Ohmic (gapless) behavior similar to that of the bare LSCO(0.35)
film shown in Fig. 1(a). However, when the thickness of the top
LSCO(0.35) layer was reduced from 10~nm to 5~nm, the differential
conductance revealed a gap in the low-energy density of states
over large parts of the sample surface, as depicted in Fig. 3. It
is possible that the STM tip is coupled to a superconducting
region at the interface [assuming that the LSCO(0.35) is in the
ballistic regime], or alternatively, that the gap is a consequence
of a proximity effect in the metallic layer due to such a region.
The latter interpretation seems more convincing in light of the
absence of coherence peaks from the bilayer data, and the fact
that the zero-bias conductance is rather high, about 75\% of its
normal state value. This should be compared with the spectra
measured on the bare LSCO(0.10) film, shown in the inset of Fig.
3, where the normalized zero-bias conductance is about 3 times
smaller and the coherence peaks are well developed. Regardless of
the mechanism responsible for the appearance of the gap, this
behavior further suggest that the $T_c$ enhancement effect does
not occur in the bulk of LSCO(0.35) layer, but is apparently
confined to the interface region. We note that
superconductivity in metal-insulator LSCO multilayers was also reported in Ref. \onlinecite{Bozovic}, yet the
doping dependence and corresponding theoretical implications were
not addressed.

Superconductivity in a two-dimensional system disappears via a
Berezinskii-Kosterlitz-Thouless (BKT) transition \cite{B,KT},
where it is destroyed by phase fluctuations due to the unbinding
of thermally-excited vortex-antivortex pairs. Consequently, we
have looked for the tell-tale signatures of a BKT transition in
our data, and found them exclusively in bilayers showing
enhancement of $T_c$, as demonstrated in Fig. 4 for the
LSCO(0.35)$\,$--$\,$LSCO(0.12) bilayer. Specifically, we have
fitted the measured temperature-dependent resistance to the
predicted BKT form $R(T) = R_0\exp(-bt^{-1/2})$, valid just above
the transition temperature $T_{BKT}$. Here $R_0$ and $b$ are
material parameters and $t=T/T_{BKT}-1$. The best fit yields
$T_{BKT} \cong 32.2K$, slightly below the value extracted from the
resistance derivative, $T_{BKT} \cong 32.6 K$, as shown in Fig. 4(a). We note that the fit is in very good agreement
with data in the temperature range of the transition. At higher
temperatures the fit deviates from the data since the resistance
of the LSCO($x$) layer exceeds that of the LSCO(0.35) layer and
the current flows primarily through the latter. The $V(I)$
characteristics are consistent with a BKT transition as well, where one expects $V\propto I^a$, with
$a=3$ just below $T_{BKT}$ and growing with decreasing
temperature. Fig. 4(b) exhibits such a behavior and provides the
estimate $T_{BKT}\cong 32.5K$, close to the values stated above.
Such signatures, indicative of a BKT transition, were not observed
for the LSCO(0.35)$\,$--$\,$LSCO(0.18) bilayer (that did not
exhibit a $T_c$ enhancement), nor on the LSCO bare films.

\begin{figure}
\includegraphics[width=\linewidth]{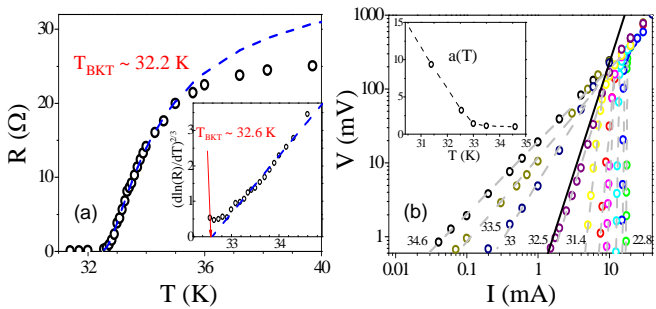}
\caption{(color online). (a) $R(T)$ of the bilayer with $x = 0.12$.
The blue dashed line is a fit to the expected BKT behavior near
the transition, yielding the estimate $T_{BKT} \approx 32.2~K$.
Inset: The same data plotted as $(\frac{dlnR}{dT})^{-\frac{2}{3}}$
versus $T$, which fits to $T_{BKT} \approx 32.6~K$.
(b) $V(I)$ characteristics at various temperatures. The solid
line corresponds to $V \propto I^3$.
The inset shows the exponent $a$ in $V\propto I^a$ as a
function of $T$.}\label{fig4}
\end{figure}

What is the reason for the enhancement? Previous reports of $T_c$
enhancement in LSCO thin films, attributed the effect either to
epitaxial compressive strain exerted by the substrate
\cite{Sato,Locquet,Sato2}, or to excess oxygenation of the film
\cite{Sato2,Bozovic-O}. Our samples were annealed in standard
oxygen environment at moderate temperatures which generally yield
a stoichiometric oxygen content \cite{Sato2}, thus making it
highly unlikely that over-oxygenation plays a role in the
enhancement reported here. The effect of compressive strain is
depicted in Fig. 2(b), where we plot $T_c$ data \cite{Sato2}, for
LSCO films grown on LaSrAlO$_4$, whose lattice constant mismatch
with our LSCO($x$) layers is somewhat larger than that of
LSCO(0.35) \cite{Locquet}. Apparently, compressive strain
increases $T_c$ for every $x$ within the superconducting region of
the phase diagram. Moreover, the original dome structure of this
region is preserved, and in particular, maintains its maximum at
$x=0.15$. The $T_c$ enhancement in our bilayers presents a
markedly different behavior, as seen in Fig. 2(a). First, it
occurs only for UD bilayers. Secondly, the original peak in $T_c$
is shifted from $x=0.15$ to the vicinity of $x=0.12$, where a dip
or flattening occurs in the $T_c$ curve of the bare films. Thus,
strain alone cannot account for the enhancement found in the
bilayer systems. Finally, since the maximal enhanced $T_c$ is far
larger than the optimal $T_c$ of the bare films, we can rule out
migration of cations across the interface as the source of the
effect.

A previous study \cite{Merchant} of an
analogous system to the bilayers discussed here, may shed light on
our findings. There, $T_c$ of a granular Pb film covered by a
silver overlayer, was found to initially increase with Ag thickness. Despite being insulating,
tunneling into the bare lead film demonstrated well-developed
superconductivity on each grain below the bulk $T_c$ of lead. Strong
phase fluctuations between the grains denied the system of
establishing global superconductivity. Apparently, the silver
enhanced the inter-grain Josephson coupling, leading to a larger
phase stiffness and higher $T_c$. The
parallels with our bilayers are compelling. Like in the granular
lead film, $T_c$ of UD cuprates is governed by their small
superfluid stiffness, while there are indications for pairing above
$T_c$ (the analogy may go even further in view of the evidence for
electronic inhomogeneities in these systems \cite{Lang}). We suggest
that pair tunneling through the metallic LSCO(0.35) overlayer
strengthens the phase coupling between locally superconducting
regions of the LSCO($x$) layer in the vicinity of the interface,
thereby enhancing $T_c$ in this portion of the sample. Such coupling
is possible since the coherence length in the LSCO(0.35) layer, at the
relevant temperatures (estimated from data presented in Ref.
\onlinecite{Tabata}), is larger than the typical spatial scale,
$\sim 2-3$ nm, of the superconducting-gap inhomogeneities in
the cuprates \cite{Lang,ourreview}. When the bottom layer is overdoped,
phase stiffness ceases to be a limiting factor and the enhancement
disappears. On the other hand, the decrease in the enhancement towards
the UD boundary of the superconducting region may reflect the reduction
of the excitation gap in this limit, as measured by angle resolved
photoemission spectroscopy (ARPES) \cite{LBCO}, and by STM \cite{Yuli}.

In view of this proposed scenario we need to recall that no
enhancement of $T_c$ was observed in our Au$\,$--$\,$LSCO(0.10)
bilayer. Such a negative result may stem from the differences in
both the Fermi wavevectors and lattice structures of the two
layers, which could significantly reduce the tunneling amplitude
through the interface. Additionally, since the induced phase
couplings in the bottom layer depend on the pair propagation
amplitude through the top metallic film, it is possible that
vestiges of pairing in the LSCO(0.35) layer play a role in
establishing the enhancement in the LSCO(0.35)$\,$--$\,$LSCO($x$)
systems. Finally, we note that the lack of enhancement in the
Au$\,$--$\,$LSCO(0.10) sample implies that screening due to the
top metallic layer is not responsible for the effect which we
measure, in contrast to Ref. \onlinecite{Aviad}.

Another distinctive feature of our data deserves attention. The
maximal enhanced $T_c$ is achieved when the UD layer is
approximately 1/8 doped. At the same doping level the lanthanum
based cuprates exhibit an anomaly in the $T_c(x)$ curve, ranging
from a local plateau, in the case of LSCO, to a substantial dip for
\LBCO \cite{LBCO1/8}. This anomaly is commonly associated with the
formation of robust static charge and spin stripe-order
\cite{la-str-Tc}. While there are many theoretical indications that
the confinement of strongly-interacting electrons to
quasi-one-dimensional systems, typically leads to a large pairing
scale \cite{ourreview}, it is also clear that such confinement
severely hampers the emergence of global phase coherence, and
consequently lowers $T_c$. The notion that pairing attains a maximum
at $x=1/8$, together with a concomitant increase in phase
fluctuations, gains support from experimental signatures as well.
Specifically, ARPES experiments show that the single-particle gap in
\LBCO is largest for $x=0.125$ \cite{LBCO}, and measurements of the
vortex-Nernst effect, which is indicative of a phase-disordered
superconducting state, find that in LSCO the maximal signal is
attained at the same doping level \cite{Ong-Nernst}. In light of
these facts it appears that the LSCO($0.35$)$\;$--$\;$LSCO(0.12)
bilayer is a system that takes advantage of the significant pairing
correlations of the $x=1/8$ state, possibly due to stripes, while
avoiding its limitations vis-\`{a}-vis phase coherence by tunneling
between regions of local superconducting order (for which stripes
are natural candidates) through the OD metallic layer.

In conclusion, we have found that the deposition of a thin,
heavily OD (metallic) LSCO film on top of an UD LSCO layer can
enhance its $T_c$ by up to 50\%, and presented evidence that the
effect takes place at the interface between the UD and OD
components. The enhancement does not occur when the bottom layer
is OD. Our findings corroborate the thesis that superconductivity
in the UD cuprates is controlled by the small phase stiffness in
this regime. The fact that the maximal enhanced $T_c$ occurs near
$x=1/8$ indicates that the optimal doping level, $x=0.15$, in bare
LSCO samples, may be a result of a suppression of the original
peak at $x=1/8$ due to phase fluctuations. It may also reflect on
the role of charge inhomogeneities (such as stripes) in these
systems, and demonstrates that once their suppressing effect on
the phase coherence is alleviated, the predicted large pairing
scale which they induce could increase $T_c$.

The authors are grateful to G. Deutscher and S. Kivelson for helpful discussions,
and to I. Felner for performing the SQUID measurements. This research
was supported by the Israel Science Foundation, the US - Israel
Binational Science Foundation (Grant No. 2004162), the Heinrich
Hertz Minerva Center for HTSC, and by the Fund for the
Promotion of Research at the Technion.

\end{document}